\documentclass[prl,aps,twocolumn,
]{revtex4}
\usepackage{amsmath,amssymb,graphicx}
\usepackage{pxfonts}
\usepackage{graphicx}
\usepackage{color}

\begin{document}

\date{\today}

\title{Transport of indirect excitons and exciton mediated spin transport in a van der Waals heterostructure in magnetic fields}

\author{Zhiwen~Zhou}
\affiliation{Department of Physics, University of California San Diego, La Jolla, CA, USA}
\author{W.~J.~Brunner}
\affiliation{Department of Physics, University of California San Diego, La Jolla, CA, USA}
\author{E.~A.~Szwed}
\affiliation{Department of Physics, University of California San Diego, La Jolla, CA, USA}
\author{L.~H.~Fowler-Gerace}
\affiliation{Department of Physics, University of California San Diego, La Jolla, CA, USA}
\author{L.~V.~Butov} 
\affiliation{Department of Physics, University of California San Diego, La Jolla, CA, USA}

\begin{abstract}
We studied transport of indirect excitons (IXs) and IX mediated spin transport in a MoSe$_2$/WSe$_2$ van der Waals heterostructure in magnetic fields up to 8~T. We observed the long-range IX transport and the long-range IX mediated spin transport in the magnetic fields. The IX transport and spin transport are characterized by the $1/e$ decay distances reaching $\sim 100$ micrometers. The decay distance of the spin transport correlates with the decay distance of IX transport. These decay distances first increase and then decrease with increasing IX density for all studied magnetic fields. The long-range IX transport and the long-range spin transport in the magnetic fields are consistent with the similar long-range transport in zero magnetic field.
\end{abstract}
\maketitle

\section{Introduction}

Spatially indirect excitons (IXs), also known as interlayer excitons, are composed from electrons and holes in separated layers in a heterostructure (HS)~\cite{Lozovik1976}. Due to the separation between the electron and hole layers, IX lifetimes exceed lifetimes of regular spatially direct excitons (DXs) by orders of magnitude~\cite{Zrenner1992}. The long lifetimes, enable IX transport over long distances and also allow IXs to form the Bose-Einstein condensate~\cite{High2012}. 

In addition to the long-range exciton transport, IXs also enable long-range spin transport. Scattering of particles carrying the spin states can cause the spin relaxation that limits the spin transfer~\cite{Dyakonov2008}. Therefore, the suppression of scattering in IX condensate can suppress the spin relaxation caused by scattering. In addition, the electron-hole separation in IXs suppresses the spin relaxation caused by electron-hole exchange~\cite{Maialle1993}. Therefore, traveling IXs can efficiently transfer spin states, allowing the realization of long-range spin transport.

IX transport is studied in various materials, including GaAs HS~\cite{Hagn1995, Larionov2000, Butov2002, Voros2005, Ivanov2006, Gartner2006, High2008, Remeika2009, Vogele2009, Lazic2010, Alloing2012, Remeika2012, Lazic2014, Alloing2014, Gorbunov2016, Leonard2021}, GaN HS~\cite{Chiaruttini2019}, and ZnO HS~\cite{Kuznetsova2015}. IX mediated spin transport in GaAs HS is also explored~\cite{Leonard2009, High2013, Violante2015, Finkelstein2017, Leonard2018}. Recently, studies of IX transport  and IX mediated spin transport were started in van der Waals HS composed of atomically thin layers of transition metal dichalcogenides (TMD). TMD HS offer a unique materials platform for studying exciton transport and exciton mediated spin transport. Both DXs in TMD HS~\cite{Ye2014, Chernikov2014, Goryca2019} and IXs in TMD HS~\cite{Fogler2014, Deilmann2018} have high binding energies reaching hundreds of meV, significantly higher than IX binding energies in HS formed from III-V or II-VI semiconductors that reach $3-4$~meV in GaAs/AlGaAs HS~\cite{Sivalertporn2012, Szwed2024}, 10~meV in AlAs/GaAs HS~\cite{Zrenner1992}, and 30~meV in ZnO HS~\cite{Morhain2005}. Due to the high binding energies, IXs in TMD HS are stable at room temperature. Furthermore, the superfluidity temperature, which can be achieved with excitons, is proportional to the exciton binding energy and the high IX binding energies in TMD HS give an opportunity to realize high-temperature superfluidity~\cite{Fogler2014}. IXs in periodic moir{\'e} potentials in TMD HS also allow exploring the Bose-Hubbard physics~\cite{Wu2018, Yu2018, Wu2017, Yu2017, Zhang2017a, Zhang2018, Ciarrocchi2019, Seyler2019, Tran2019, Jin2019, Alexeev2019, Jin2019a, Shimazaki2020, Wilson2021, Gu2022}. 

DX transport~\cite{Kumar2014, Kulig2018, Cadiz2018, Leon2018, Leon2019, Hao2020, Datta2022}, IX transport~\cite{Rivera2016, Jauregui2019, Unuchek2019a, Unuchek2019, Liu2019, Choi2020, Huang2020, Yuan2020, Li2021, Wang2021, Shanks2022, Sun2022, Tagarelli2023, Rossi2023, Gao2024, Zhang2024, Fowler-Gerace2021, Peng2022, Wietek2024, Troue2023, Cutshall2025}, DX mediated spin transport~\cite{Onga2017}, and IX mediated spin transport~\cite{Rivera2016, Unuchek2019, Huang2020, Shanks2022} are intensively studied in TMD materials. (Due to the coupling of the spin and valley indices~\cite{Xiao2012, Cao2012, Zeng2012, Mak2012} the spin transport is coupled to the valley transport in TMD HS and, therefore, for simplicity, we will use the term 'spin' also for 'spin-valley'.) These studies showed that in-plane disorder suppress diffusive IX transport and IX mediated spin transport due to IX localization and scattering: In TMD HS, even in the case of long IX lifetimes, diffusive IX transport~\cite{Rivera2016, Jauregui2019, Unuchek2019a, Unuchek2019, Liu2019, Choi2020, Huang2020, Yuan2020, Li2021, Wang2021, Shanks2022, Sun2022, Tagarelli2023, Rossi2023, Gao2024, Zhang2024, Wietek2024} and IX mediated spin transport~\cite{Rivera2016, Unuchek2019, Huang2020, Shanks2022} are characterized by low $1/e$ decay distances $d_{1/e}$ and $d^{\rm s}_{1/e}$, respectively, up to a few micrometers. Recent studies showed that long-range IX transport and long-range IX mediated spin transport with $d_{1/e}$ and $d^{\rm s}_{1/e}$ reaching 100~micrometers can be realized in TMD HS~\cite{Fowler-Gerace2024, Zhou2024}.

In this work, we studied IX transport and IX mediated spin transport in a MoSe$_2$/WSe$_2$ van der Waals HS in magnetic fields up to 8~T. The dispersion relation of excitons in high magnetic fields was theoretically studied for three-dimensional (3D) excitons~\cite{Gorkov1968}, 2D excitons in a perpendicular magnetic field~\cite{Lerner1980, Kallin1984, Paquet1985}, and IXs in a perpendicular magnetic field~\cite{Lozovik1997}. It was shown that high magnetic fields induce coupling between the exciton internal structure and c.m. motion~\cite{Gorkov1968, Lerner1980, Kallin1984, Paquet1985, Lozovik1997}. This coupling modifies the exciton dispersion and enhances the exciton mass that was measured for IXs in GaAs/AlGaAs HS in Refs.~\cite{Butov2001, Lozovik2002}. The studies of IX transport in GaAs/AlGaAs HS showed a strong reduction of IX transport distance with increasing magnetic field~\cite{Kuznetsova2017, Dorow2017}. In this work, we probe the effect of magnetic field on IX transport in van der Waals TMD HS.

\begin{figure}
\begin{center}
\includegraphics[width=8.7cm]{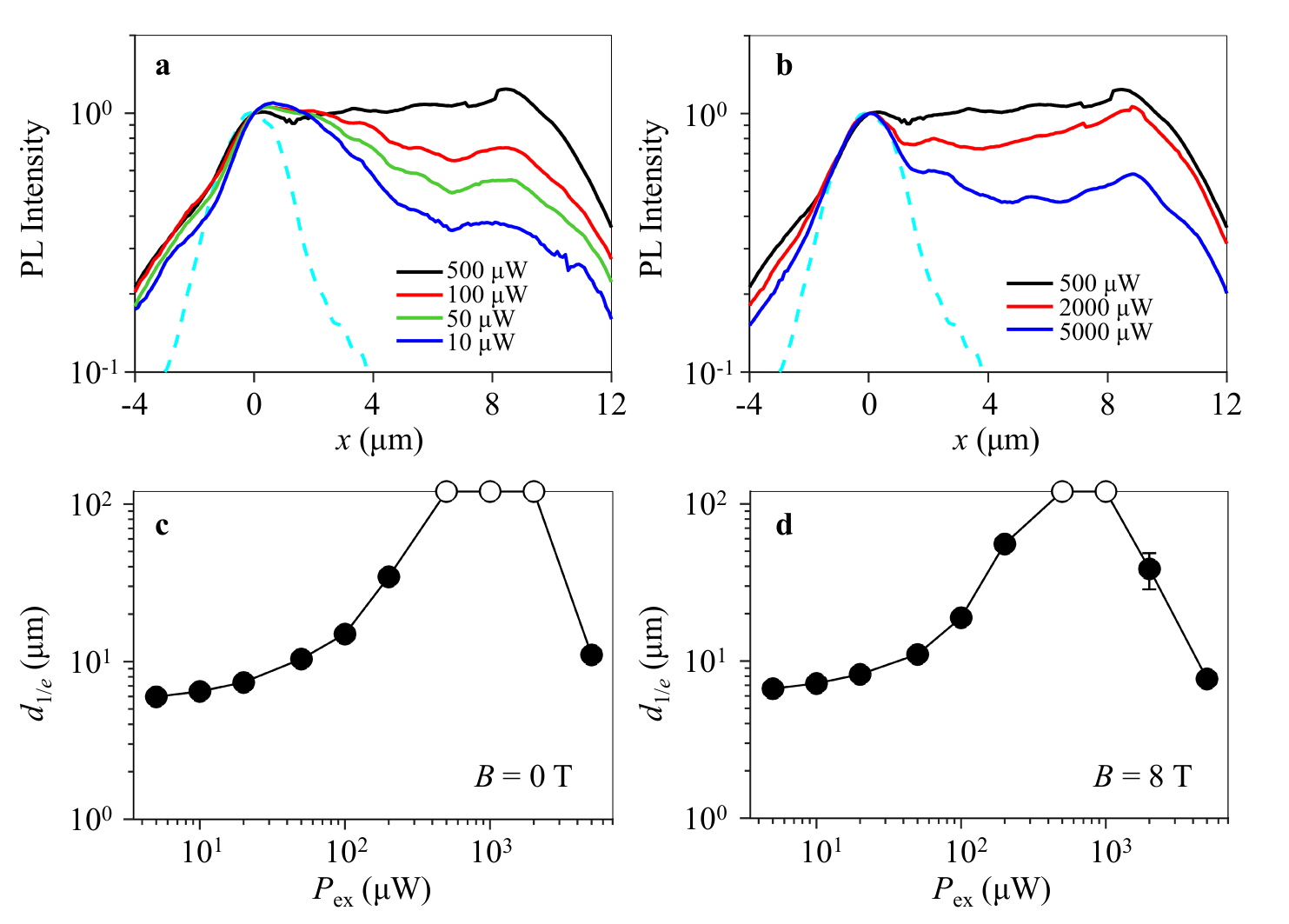}
\caption{{\bf IX transport in magnetic fields}.  
(a,b) Normalized LE-IX luminescence profiles in magnetic field $B = 8$~T for laser excitation power $P_{\rm ex} = 500$, 100, 50, 10 $\mu$W (top to bottom) in (a) and $P_{\rm ex} = 500$, 2000, 5000 $\mu$W (top to bottom) in (b). The dashed line shows the DX luminescence profile in the MoSe$_2$ monolayer; this profile is close to the laser excitation profile for short-range DX transport. The $\sim 2~\mu$m laser spot is centered at $x = 0$. (c,d) The $1/e$ decay distance of IX transport $d_{1/e}$ vs. $P_{\rm ex}$ for magnetic field $B = 0$ (c) and 8~T (d). The $d_{1/e}$ values are obtained from least-squares fitting the LE-IX luminescence profiles to exponential decays in the region $x = 0 - 9$~$\mu$m, from the excitation spot to the HS edge. The data with the fit indicating diverging $d_{1/e}$ are presented by circles on the edge.
}
\end{center}
\end{figure}

\section{Experiment}

We consider IX transport in MoSe$_2$/WSe$_2$ HS where adjacent MoSe$_2$ monolayer and WSe$_2$ monolayer form the separated electron and hole layers for IXs~\cite{Rivera2015}. The HS details are presented in Appendix A. 

IXs are generated by focused laser excitation with the laser excitation energy resonant to DX in WSe$_2$ HS layer. As shown in Ref.~\cite{Fowler-Gerace2024}, the resonant excitation enhances the IX transport distances. As in earlier studies~\cite{Rivera2016, Jauregui2019, Unuchek2019a, Unuchek2019, Liu2019, Choi2020, Huang2020, Yuan2020, Li2021, Wang2021, Shanks2022, Sun2022, Tagarelli2023, Rossi2023, Gao2024, Zhang2024, Fowler-Gerace2021, Peng2022, Wietek2024, Troue2023, Cutshall2025, Fowler-Gerace2024}, the propagation of the IX PL intensity from the laser excitation spot is measured via spatially-resolved imaging to characterize the IX transport. In turn, as in earlier studies~\cite{Rivera2016, Unuchek2019, Huang2020, Shanks2022, Zhou2024}, the propagation of the circularly polarized IX PL intensity from the laser excitation spot is measured via spatially- and polarization-resolved imaging to characterize the IX mediated spin transport. The measurement details are presented in Appendix B.

Figure~4 in Appendix C shows IX PL spectra at different laser excitation powers $P_{\rm ex}$. The spectra at zero magnetic field (Fig.~4b,e,h) are similar to the corresponding zero-field spectra considered in Ref.~\cite{Zhou2024}: A single lower-energy IX (LE-IX) line is observed in the spectra at low $P_{\rm ex}$ and an additional higher-energy IX (HE-IX) line appears in the spectra at high $P_{\rm ex}$. The HE-IX line was attributed to the appearance of moir{\'e} cells with double occupancy in Ref.~\cite{Zhou2024}. Figure~4 shows that similar IX spectra are observed in the studied magnetic fields $-8$~T~$< B < 8$~T. In this work, we consider the propagation of the LE-IX PL intensity from the laser excitation spot to characterize the IX transport (Figs.~1-3). The LE-IX spectra are separated from the HE-IX spectra by the Gaussian fits (Fig.~4). Figure~5 in Appendix~D shows that considering the spectrally integrated IX PL intensity gives qualitatively similar IX transport characteristics. 

Figures~1a,b show LE-IX PL profiles $I(x)$ in magnetic field $B = 8$~T for different laser excitation power $P_{\rm ex}$. IXs propagate from the laser excitation spot centered at $x = 0$. The excitation spot is presented in Figs.~1a,b by the DX luminescence profile in the MoSe$_2$ monolayer; this profile is close to the laser excitation profile for short-range DX transport. The IX transport from the excitation spot is quantified by the $1/e$ decay distance of LE-IX PL $d_{1/e}$. The $d_{1/e}$ values are obtained from least-squares fitting the LE-IX luminescence profiles $I(x)$ to exponential decays in the region $x = 0 - 9$~$ \mu$m, from the excitation spot to the HS edge. 

For zero magnetic field, the IX transport enhances with increasing $P_{\rm ex}$ for $P_{\rm ex} \lesssim 500$~$\mu$W and reduces with increasing $P_{\rm ex}$ for higher $P_{\rm ex}$ as shown by the variation of $d_{1/e}$ with $P_{\rm ex}$ in Fig.~1c. A similar dependence of IX transport on the excitation density was observed in earlier studies in zero magnetic field in Ref.~\cite{Fowler-Gerace2024}. 

IX transport in magnetic fields also non-monotonically depends on density (Fig.~1a,b), with $d_{1/e}$ first increasing and then reducing with $P_{\rm ex}$ (Fig.~1d). Furthermore, the long-range IX transport with $d_{1/e}$ reaching 100~$\mu$m is realized in magnetic fields (Fig.~1d). 

IX signal vanishes outside HS and, for the fit of the data in the HS range $x = 0 - 9$~$\mu$m, from the excitation spot to the HS edge, the uncertainties in the fit allow determining $d_{1/e}$ up to $\sim 100$~$\mu$m: For $d_{1/e} \lesssim 100$~$\mu$m, the uncertainties in the fit (shown by the error bars when larger than the point size in Fig.~1c,d) are small; however, for $d_{1/e}$ extracted from the fit higher than $\sim 100$~$\mu$m, the uncertainties become large and comparable to the extracted $d_{1/e}$.

\begin{figure}
\begin{center}
\includegraphics[width=8.7cm]{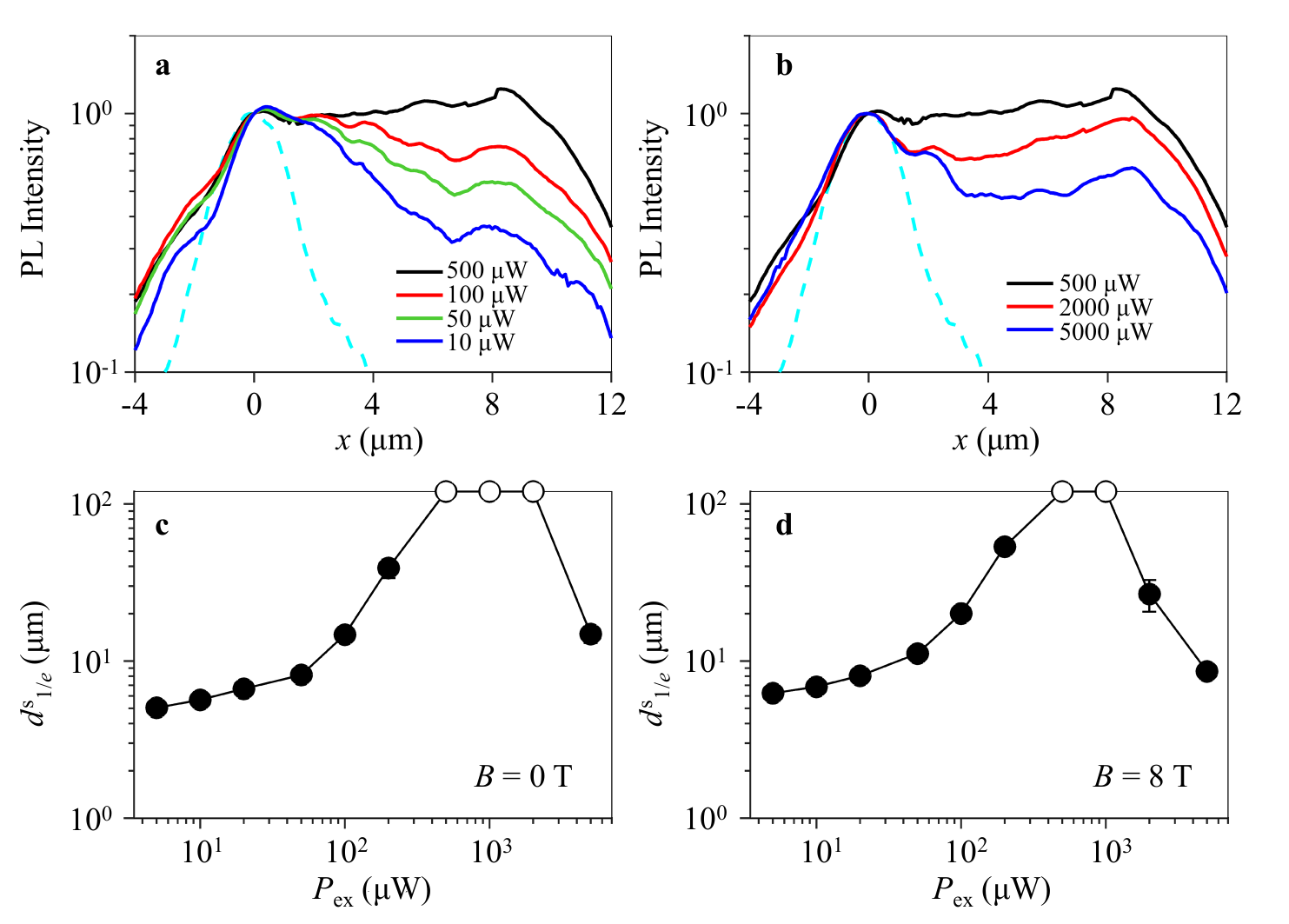}
\caption{{\bf IX mediated spin transport in magnetic fields}. 
(a,b) Normalized LE-IX spin density profiles $I_{\rm spin} = I_{\sigma^+} - I_{\sigma^-}$ in magnetic field $B = 8$~T for laser excitation power $P_{\rm ex} = 500$, 100, 50, 10 $\mu$W (top to bottom) in (a) and $P_{\rm ex} = 500$, 2000, 5000 $\mu$W (top to bottom) in (b). The dashed line shows the DX luminescence profile in the MoSe$_2$ monolayer; this profile is close to the laser excitation profile for short-range DX transport. The $\sim 2~\mu$m laser spot is centered at $x = 0$. (c,d) The $1/e$ decay distance of IX mediated spin density transport $d^{\rm s}_{1/e}$ vs. $P_{\rm ex}$ for magnetic field $B = 0$ (c) and 8~T (d). The $d^{\rm s}_{1/e}$ values are obtained from least-squares fitting the $I_{\rm spin}(x)$ profiles to exponential decays in the region $x = 0 - 9$~$\mu$m, from the excitation spot to the HS edge. The data with the fit indicating diverging $d^{\rm s}_{1/e}$ are presented by circles on the edge.
}
\end{center}
\end{figure}

IX transport considered above is characterized by the propagation of total IX PL intensity in both circular polarizations $n \sim I_{\sigma^+} + I_{\sigma^-}$, where $I_{\sigma^+}$ is the intensity of IX PL co-polarized with the circularly polarized laser excitation and $I_{\sigma^-}$ is the intensity of cross-polarized IX PL. The degree of circular polarization of IX PL $P = (I_{\sigma^+} - I_{\sigma^-})/(I_{\sigma^+} + I_{\sigma^-})$. The transport of spin polarization density carried by IXs is characterized by the propagation of $I_{\rm spin} = Pn = I_{\sigma^+} - I_{\sigma^-}$. Figures~2a,b show LE-IX spin density profiles $I_{\rm spin}(x)$ in magnetic field $B = 8$~T for different $P_{\rm ex}$. The IX mediated spin transport from the excitation spot is quantified by the $1/e$ decay distance of spin density profiles $d^{\rm s}_{1/e}$. The $d^{\rm s}_{1/e}$ values are obtained from least-squares fitting the $I_{\rm spin}(x)$ profiles to exponential decays in the region $x = 0 - 9$~$\mu$m, from the excitation spot to the HS edge.

For zero magnetic field, the IX mediated spin transport enhances with increasing $P_{\rm ex}$ for $P_{\rm ex} \lesssim 500$~$\mu$W and reduces with increasing $P_{\rm ex}$ for higher $P_{\rm ex}$ as shown by the variation of $d^{\rm s}_{1/e}$ with $P_{\rm ex}$ in Fig.~2c. A similar dependence of IX mediated spin transport on the excitation density was observed in earlier studies in zero magnetic field in Ref.~\cite{Zhou2024}.

\begin{figure*}
\begin{center}
\includegraphics[width=18cm]{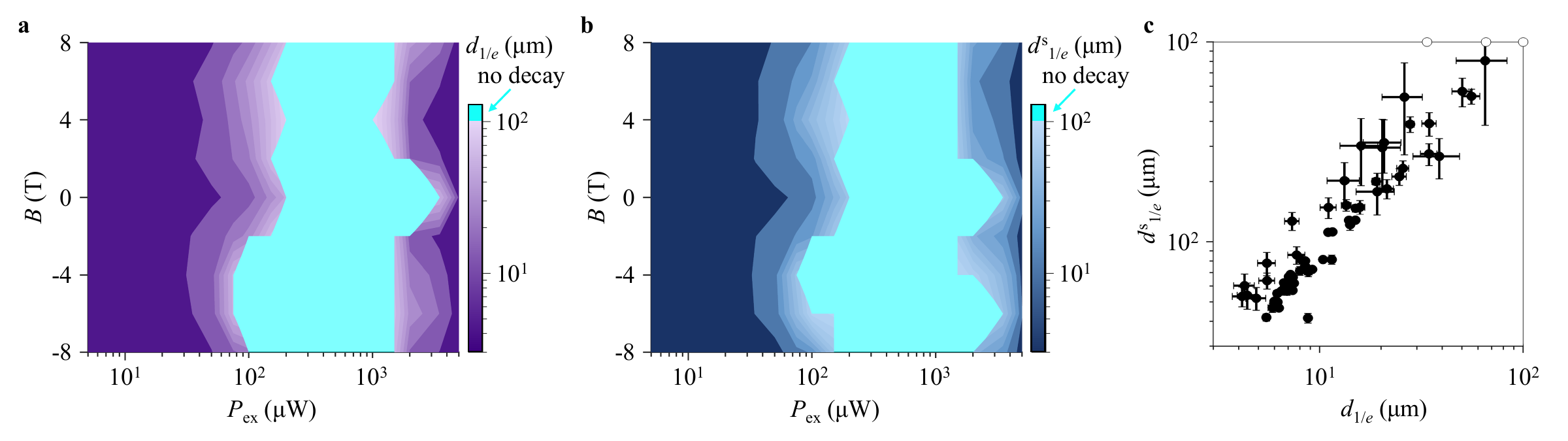}
\caption{{\bf Excitation power and magnetic field dependence of IX transport and IX mediated spin transport}.
(a-b) The $1/e$ decay distance of LE-IX transport $d_{1/e}$ (a) and the $1/e$ decay distance of LE-IX mediated spin density transport $d^{\rm s}_{1/e}$ (b) vs. $P_{\rm ex}$ and $B$. The $d_{1/e}$ and $d^{\rm s}_{1/e}$ values are obtained from least-squares fitting the LE-IX luminescence profiles and $I_{\rm spin}$ profiles, respectively, to exponential decays in the region $x = 0 - 9$~$\mu$m, from the excitation spot to the HS edge. The data with the fit indicating diverging $d_{1/e}$ (a) or $d^{\rm s}_{1/e}$ (b) are presented by cyan color. (c) Correlation between $d^{\rm s}_{1/e}$ and $d_{1/e}$. The values for $d_{1/e}$ and $d^{\rm s}_{1/e}$ are taken from (a) and (b). The data with the fit indicating diverging $d^{\rm s}_{1/e}$ and $d_{1/e}$ are presented by circles on the edge. The error bars represent the uncertainty in least-squares fitting the transport decays to exponential decays. The enhancement of $d^{\rm s}_{1/e}$ with $d_{1/e}$ is observed in a broad range of both excitation density and magnetic field variations, corresponding to the range of these parameters in (a, b). 
}
\end{center}
\end{figure*}

IX mediated spin transport in magnetic fields also non-monotonically depends on density (Fig.~2a,b), with $d^{\rm s}_{1/e}$ first increasing and then reducing with $P_{\rm ex}$ (Fig.~2d). Furthermore, the long-range spin transport with $d^{\rm s}_{1/e}$ reaching 100~$\mu$m is realized in magnetic fields (Fig.~2d). As for the IX transport decay distance $d_{1/e}$ outlined above, the uncertainties in the fit allow determining the spin transport decay distance $d^{\rm s}_{1/e}$ up to $\sim 100$~$\mu$m. 

We measured IX transport and IX mediated spin transport for magnetic fields $-8$~T~$< B < 8$~T. For all studied magnetic fields, the IX transport decay distance $d_{1/e}$ (Fig.~3a) and spin transport decay distance $d^{\rm s}_{1/e}$ (Fig.~3b) reach $\sim 100$~$\mu$m. Both $d_{1/e}$ and $d^{\rm s}_{1/e}$ first increase and then decrease with increasing IX density for all studied magnetic fields (Fig.~3a,b). 

Figure~3c presents the correlation between $d_{1/e}$ and $d^{\rm s}_{1/e}$. The values for $d_{1/e}$ and $d^{\rm s}_{1/e}$ in Fig.~3c are taken from Fig.~3a and Fig.~3b. Figure~3c shows that the decay distance of the spin transport correlates with the decay distance of the IX transport. The enhancement of $d^{\rm s}_{1/e}$ with $d_{1/e}$ is observed in a broad range of both excitation density and magnetic field variations, corresponding to the range of these parameters in Figs.~3a,b.

\section{Discussion}

{\bf IX transport.}
For diffusive IX transport in a disordered in-plane potential in a heterostructure, the IX transport decay distance $d_{1/e}$ increases with density due to the enhancement of IX screening of the disorder potential and the enhancement of repulsive interaction between IXs~\cite{Ivanov2006, Remeika2009, Ivanov2002}. However, for the IX long-range transport, a different behavior is observed: $d_{1/e}$ drops with density at high densities (Fig.~1c). The disagreement with the diffusive IX transport is outlined in the studies of the IX long-range transport in $B = 0$ in Ref.~\cite{Fowler-Gerace2024}.

The Bose-Hubbard model predicts the superfluid phase for the number of bosons per lattice site $N \sim 1/2$ and the insulating phase (Mott insulator or the Bose glass) for $N \sim 0$ and $N \sim 1$~\cite{Fisher1989}. The IX long-range transport and the enhancement followed by the suppression of $d_{1/e}$ with density (Fig.~1d) are consistent with this prediction. For the IX long-range transport, the mean-field estimate for the IX density $n = \varepsilon \delta E / (4 \pi e^2 d_z)$~\cite{Yoshioka1990} gives $n \sim 2 \cdot 10^{11}$~cm$^{-2}$ for the measured IX energy shift in the regime of IX long-range transport $\delta E \sim 3$~meV ($d_z \sim 0.6$~nm is the separation between the electron and hole layers, $\varepsilon \sim 7.4$ is the dielectric constant~\cite{Laturia2018}). $N \sim 1/2$ for this density and the moir{\'e} superlattice period $b = 17$~nm corresponding to the twist angle $\delta \theta = 1.1^\circ$ that agrees with the angle between MoSe$_2$ and WSe$_2$ layers in the HS ($b \sim a/\delta \theta$, $a$ is the lattice constant). The agreement with the Bose-Hubbard model is outlined in the studies of the IX long-range transport in $B = 0$ in Ref.~\cite{Fowler-Gerace2024}.

In this work, we extended the IX transport measurements to magnetic fields up to 8~T. We observed the IX long-range transport with the decay distance $d_{1/e}$ reaching $\sim 100$~$\mu$m in the magnetic fields (Figs.~1d and 3a). We also observed that $d_{1/e}$ first enhances and then decreases with increasing IX density in the magnetic fields (Figs.~1d and 3a). The IX long-range transport and the non-monotonic IX transport dependence on density in the magnetic fields in this work are consistent with the similar IX long-range transport and the non-monotonic IX transport dependence on density in zero magnetic field in Ref.~\cite{Fowler-Gerace2024} and, in turn, are consistent with the Bose-Hubbard model as outlined above.

The IX transport dependences on density and on magnetic field in the MoSe$_2$/WSe$_2$ HS (Figs.~1d and 3a) are strongly different from the IX transport dependences on density and on magnetic field in GaAs/AlGaAs HS~\cite{Ivanov2006, Remeika2009, Kuznetsova2017, Dorow2017}. In contrast to the monotonic enhancement of IX transport with IX density in GaAs/AlGaAs HS both in zero magnetic field~\cite{Ivanov2006, Remeika2009} and in magnetic fields up to 10~T~\cite{Kuznetsova2017}, IX transport in the MoSe$_2$/WSe$_2$ HS non-monotonically varies with IX density in all studied magnetic fields (Figs.~1d and 3a) as outlined above. This difference is consistent with the lack of a moir{\'e} superlattice potential for IXs in GaAs/AlGaAs HS and the presence of the moir{\'e} superlattice potential for IXs in MoSe$_2$/WSe$_2$ HS: as outlined above, the transition from IX localization to IX long-range transport to IX localization with increasing IX density is in qualitative agreement with the Bose-Hubbard model prediction for superfluid and insulating phases in periodic potentials of moir{\'e} superlattices.

Furthermore, in contrast to the strong suppression of IX transport distances in GaAs/AlGaAs HS by the magnetic field~\cite{Kuznetsova2017, Dorow2017}, no such suppression is observed in MoSe$_2$/WSe$_2$ HS (Figs.~1c,d). This difference is consistent with the theory of magnetoexcitons~\cite{Gorkov1968, Lerner1980, Kallin1984, Paquet1985, Lozovik1997, Butov2001, Lozovik2002}. The IX transport in GaAs/AlGaAs HS is suppressed due to the enhancement of IX mass in the magnetic field~\cite{Kuznetsova2017, Dorow2017}. A strong enhancement of the exciton mass is realized in the high-magnetic-field regime when the cyclotron energy is comparable or larger than the exciton binding energy~\cite{Gorkov1968, Lerner1980, Kallin1984, Paquet1985, Lozovik1997, Butov2001, Lozovik2002}. Due to the small IX binding energy and the high electron and hole cyclotron energies originating from the small electron and hole masses in GaAs/AlGaAs HS, the high-magnetic field regime is realized in the magnetic fields $\sim 10$~T in GaAs/AlGaAs HS and the strong enhancement of the IX mass in these magnetic fields~\cite{Butov2001, Lozovik2002} strongly suppresses the IX transport in GaAs/AlGaAs HS~\cite{Kuznetsova2017, Dorow2017}. In contrast, due to the high IX binding energy and the low electron and hole cyclotron energies originating from the large electron and hole masses in MoSe$_2$/WSe$_2$ HS, the high-magnetic field regime is not realized in the magnetic fields $\sim 10$~T in MoSe$_2$/WSe$_2$ HS and neither strong enhancement of the IX mass nor strong suppression of the IX transport is expected in these magnetic fields in MoSe$_2$/WSe$_2$ HS, that is consistent with the data in Figs.~1c,d and 3a.

{\bf IX mediated spin transport.}
The long-range spin transport with the decay length reaching $\sim 100$~$\mu$m and the correlation of the spin transport decay length $d^{\rm s}_{1/e}$ with the IX transport decay length $d_{1/e}$ were observed in the studies of IX mediated spin transport in $B = 0$ in Ref.~\cite{Zhou2024}. In this work, we extended the measurements of IX mediated spin transport to magnetic fields up to 8~T. We observed the long-range spin transport with the decay length reaching $\sim 100$~$\mu$m in the magnetic fields (Figs.~2d, 3b). We also observed the correlation of the spin transport decay length with the IX transport decay length in the magnetic fields (Fig.~3c). The long-range spin transport and its correlation with the IX transport in the magnetic fields are consistent with the similar long-range spin transport and its correlation with the IX transport in zero magnetic field in Ref.~\cite{Zhou2024}.
The correlation of the longer-range spin transport with the longer-range IX transport and, in turn, with the suppressed IX scattering complies with the suppression of the spin relaxation due to the suppression of scattering of particles carrying the spin states~\cite{Dyakonov2008}.

\section{Conclusion}

We studied transport of IXs and IX mediated spin transport in a MoSe$_2$/WSe$_2$ heterostructure in magnetic fields up to 8~T. We observed the long-range IX transport and the long-range IX mediated spin transport with the $1/e$ decay distances reaching $\sim 100$~$\mu$m in the magnetic fields.  The decay distance of the spin transport correlates with the decay distance of IX transport. These decay distances first increase and then decrease with increasing IX density in the magnetic fields. The long-range IX transport and the long-range IX mediated spin transport in the magnetic fields are consistent with the similar long-range transport in zero magnetic field.

\section{Acknowledgments}

We thank Misha Fogler for discussions. The experiments were supported by the Department of Energy, Office of Basic Energy Sciences, under award DE-FG02-07ER46449. The heterostructure manufacturing was supported by NSF Grant 1905478. The data analysis was supported by NSF Grant 2516006.

\section{Appendix A: Heterostructure}

The MoSe$_2$/WSe$_2$ HS was assembled using the dry-transfer peel technique~\cite{Withers2015}. The HS manufacturing details are described in Ref.~\cite{Fowler-Gerace2024}. The same HS was used for studies of IX transport in Ref.~\cite{Fowler-Gerace2024} and for studies of IX mediated spin transport in Ref.~\cite{Zhou2024} in zero magnetic field. The thickness of the bottom hBN layer is $\sim 40$~nm, the thickness of the top hBN layer is $\sim 30$~nm. The MoSe$_2$ monolayer is on top of the WSe$_2$ monolayer. 
The hBN layers cover the entire areas of MoSe$_2$ and WSe$_2$ layers.
The WSe$_2$ and MoSe$_2$ edges were used for a rotational alignment between the WSe$_2$ and MoSe$_2$ monolayers. The twist angle between the monolayers $\delta \theta = 1.1^\circ$ corresponding to the moir{\'e} superlattice period $b = 17$~nm agrees with the angle between MoSe$_2$ and WSe$_2$ edges in the HS~\cite{Fowler-Gerace2024, Zhou2024}. 
The moir{\'e} potentials can be affected by atomic reconstruction~\cite{Weston2020, Rosenberger2020, Zhao2023} and by disorder and may vary over the HS area. 
The measured IX $g$-factor $\sim - 16$ corresponds to H stacking in the MoSe$_2$/WSe$_2$ HS~\cite{Seyler2019, Wozniak2020}.

\section{Appendix B: Measurements}

Excitons were generated by a cw Ti:Sapphire laser with the excitation energy $E_{\rm ex} = 1.689$~eV resonant to DX in WSe$_2$ HS layer. PL spectra were measured using a spectrometer with a resolution of 0.2~meV and a liquid-nitrogen-cooled CCD. Representative polarization-resolved IX PL spectra are presented in Appendix C. The spatial profiles of polarization-resolved IX PL vs. $x$ were obtained from the polarization-resolved PL images detected using the CCD with the spatial resolution 0.8~$\mu$m. The signal was integrated within 1~$\mu$m in $y$ direction. 
For a direct comparison with earlier measurements in zero magnetic field in Refs.~\cite{Fowler-Gerace2024, Zhou2024}, IX transport and IX mediated spin transport in magnetic fields were studied along the same path.

The measurements were performed in an optical dilution refrigerator at temperature 4~K and magnetic fields up to 8~T perpendicular to the HS layer plane. The sample was mounted on an Attocube $xyz$ piezo translation stage allowing adjusting the sample position relative to a focusing lens installed inside the fridge.

\section{Appendix C: IX PL spectra}

\begin{figure}
\begin{center}
\includegraphics[width=8cm]{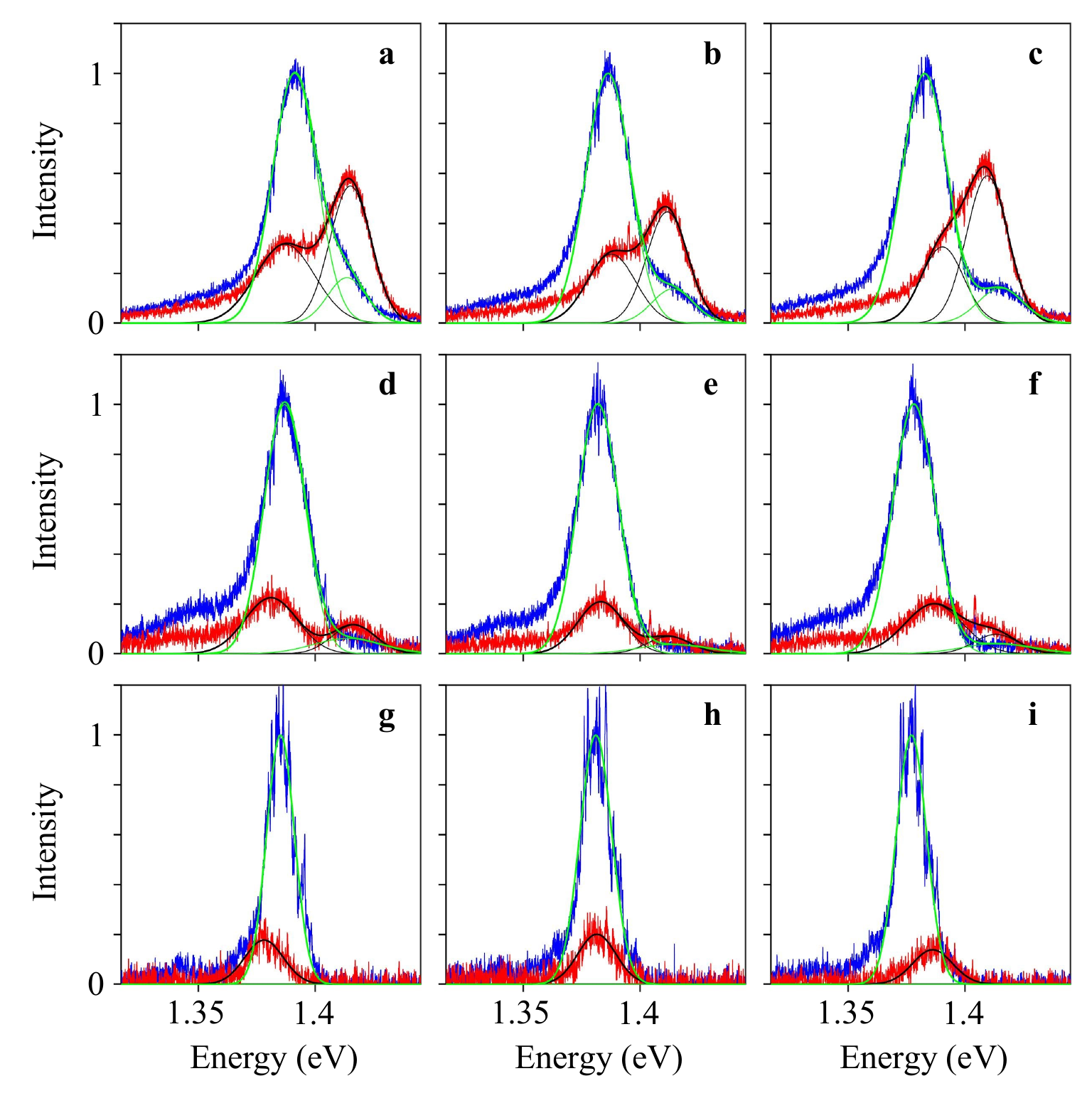}
\caption{{\bf IX PL spectra.} 
Co-polarized (blue) and cross-polarized (red) IX spectra for the excitation power $P_{\rm ex} = 5$~mW (a,b,c),  500~mW (d,e,f), and 5~$\mu$W (g,h,i) in magnetic field $B = - 8$~T (a,d,g), 0 (b,e,h), and 8~T (c,f,i). The lower-energy IX (LE-IX) PL is co-polarized. The higher-energy IX (HE-IX) PL is cross-polarized. The HE-IXs appear in the spectra at high $P_{\rm ex}$.
PL spectra are normalized to the maximum of the co-polarized intensity.
The Gaussian fits to the co-polarized (cross-polarized) LE-IX spectra and HE-IX spectra are shown by the thin green (black) lines. The sum of the Gaussians shown by the thin green (black) lines is
shown by the thick green (black) line.
}
\end{center}
\end{figure}

Figure~4 shows IX PL spectra at different laser excitation powers $P_{\rm ex}$ and magnetic fields $B$. The spectra at $B=0$ (Fig.~4b,e,h) are similar to the corresponding zero-field spectra in Ref.~\cite{Zhou2024}: A single LE-IX PL line is observed in the spectra at low $P_{\rm ex}$ and an additional HE-IX PL line appears in the spectra at high $P_{\rm ex}$. Similar IX spectra are observed in the magnetic fields $-8$~T~$< B < 8$~T as shown for $B = - 8$~T in Fig.~4a,d,g and for $B =  8$~T in Fig.~4c,f,i.

\section{Appendix D: Excitation power and magnetic field dependence of IX transport}

In the main text, we consider the propagation of the LE-IX PL intensity from the laser excitation spot to characterize the IX transport (Figs.~1-3). The LE-IX spectra are separated from the HE-IX spectra by the Gaussian fits as shown in Fig.~4. Figure~5 shows that considering the spectrally integrated IX PL intensity including both LE-IX and HE-IX gives qualitatively similar IX transport characteristics (compare Fig.~3a for LE-IX and Fig.~5 for the spectrally integrated IX PL intensity including both LE-IX and HE-IX).

\begin{figure}
\begin{center}
\includegraphics[width=7cm]{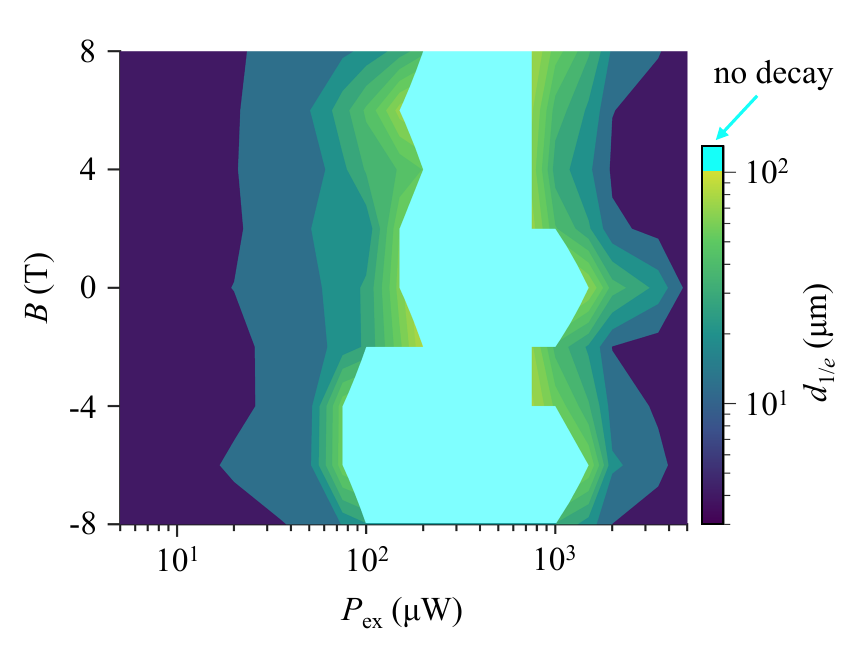}
\caption{{\bf Excitation power and magnetic field dependence of IX transport}. 
The $1/e$ decay distance of spectrally integrated IX luminescence $d_{1/e}$ vs. $P_{\rm ex}$ and $B$. The $d_{1/e}$ values are obtained from least-squares fitting the spectrally integrated IX luminescence profiles to exponential decays in the region $x = 0 - 9~\mu$m, from the excitation spot to the HS edge. The data with the fit indicating diverging $d_{1/e}$ are presented by cyan color. Figure~5 is similar to Fig.~3a, however, Fig.~5 shows $d_{1/e}$ for the spectrally integrated IX PL including LE-IX and HE-IX, and Fig.~3a shows $d_{1/e}$ for LE-IX.
}
\end{center}
\end{figure}

\end{document}